\documentclass[final,3p,times,twocolumn]{elsarticle}
 \biboptions{comma,sort&compress}
\usepackage{here}
 \usepackage{graphicx}
  \usepackage{epsfig}

\def\nin{\noindent}
\def\beq{\begin{equation}}
\def\eeq{\end{equation}}
\def\bea{\begin{eqnarray}}
\def\eea{\end{eqnarray}}

\newcommand{\ba}{\begin{array}{c}}
\newcommand{\ea}{\end{array}}
\newcommand{\nn}{\nonumber}

\newcommand{\be}{\begin{equation}}
\newcommand{\ee}{\end{equation}}

\newcommand{\ket}{\,\rangle}
\newcommand{\bra}{\langle \,}

\newcommand{\mL}{\mathcal{L}}

\newcommand{\Frac}[2]{\frac{\displaystyle #1}{\displaystyle #2}}
\newcommand{\cO}{{\cal O}}

\journal{Nuc. Phys. (Proc. Suppl.)}

\begin{document}

\begin{frontmatter}



\title{Scalar and pseudoscalar correlators in Resonance Chiral Theory}

 \author[label1]{J.J.  Sanz-Cillero\corref{cor1}}
  \address[label1]{Grup de F\'\i sica Te\`orica and IFAE, Universitat
Aut\`onoma de Barcelona,\\ E-08193 Barcelona, Spain}
\cortext[cor1]{Speaker}
\ead{cillero@ifae.es}



\begin{abstract}
\noindent
The SU(3) octet $SS-PP$ correlator
and the difference of the singlet and octet scalar correlators
are computed  within Resonance
Chiral Theory.
The calculation is carried on
up to the one-loop level, i.e., up to next-to-leading order
in the $1/N_C$ expansion.
Using the resonance expressions as interpolators between
long and short distances,
we    demand the correlators to  follow
the high-energy power behavior prescribed
by the operator product expansion
and extract  predictions for the low-energy constants.
By adding more and more complicated
operators to the  hadronic action,  our
description  is  progressively improved,
producing for the SS-PP correlator
the chiral coupling   estimates
$L_8(\mu)  = (1.0\pm 0.4)\cdot 10^{-3}$ and
$C_{38}(\mu)  = (8\pm 5)\cdot 10^{-6}$ for $\mu=770$~MeV.
Some first results  for the $S_1S_1-S_8S_8$ correlator are shown,
like, for instance, the positivity of
the spectral function for
two--meson cuts, the high-energy constraints for each
separate scalar form-factor   and  preliminary expressions
for the $L_6(\mu)$ low-energy constant.
\end{abstract}

\begin{keyword}

Chiral Lagrangians, $1/N_C$ expansion

\end{keyword}

\end{frontmatter}


\section{Introduction}
\nin

In this talk~\footnote{
Talk given at 
QCD'10 (25th anniversary)
--15th International QCD Conference--, 
28th June - 3rd July 2010 Montpellier (France).     
This work
was supported in part by
the Ministerio de Ciencia e Innovaci´on
under grant CICYTFEDER-FPA2008-01430,
the Juan de la Cierva Program,
the EU Contract No. MRTN-CT-2006-035482 --"FLAVIAnet"--,
the Spanish Consolider-Ingenio 2010 Programme CPAN (CSD2007-00042)
and the Generalitat de Catalunya under grant SGR2009-00894.
}
 we discuss the results from a previous work on SS-PP
correlator~\cite{L8-Trnka}
together with some new estimates
for the OZI suppressed difference of the
SU(3) singlet and octet scalar correlators
($S_1S_1-S_8S_8$)~\cite{L6-Trnka}.
We will work with a chiral invariant action for
the chiral pseudo-Goldstones  and the meson resonances,
namely, resonance chiral theory (R$\chi$T)~\cite{RChTa,RChTb}.
The large--$N_C$ limit and the $1/N_C$ expansion are taken as guiding
principles to sort out the different contributions~\cite{NC}, being the
leading order (LO) provided by the tree-level diagrams and the
meson loops suppressed by $1/N_C$~\cite{NC,Cillero-thesis,Rosell-thesis}.
This framework ensures the proper recovery of
chiral perturbation theory ($\chi$PT) at long
distances~\cite{chpt}.

The procedure followed in the analysis of both correlators
is identical. First, the  structure of the
amplitude is constrained by imposing the power behaviour prescribed
by the operator product expansion (OPE)
for deep Euclidean momentum $p^2\to-\infty$~\cite{SVZ}.
After that, one is ready to go to the low-energy regime
to make predictions on the  $\chi$PT  couplings~\cite{chpt},
using the R$\chi$T expression as an interpolator between
short and long distances~\cite{L8-Trnka}.
The phenomenological determinations for the
LEC $L_8(\mu)$ and $C_{38}(\mu)$, which rule the SS-PP correlator
at long distances,  were found in good agreement with previous
results in the bibliography~\cite{chpt,L8-Pich,L8-others}.

This same machinery has been  also applied to the $S_1S_1-S_8S_8$
correlator.  It is governed at low energies by the
$L_6(\mu)$ low-energy constant (LEC), which happens
to be suppressed in the    large--$N_C$ limit~\cite{NC}.
The study of this amplitude casts some first interesting results on the
relation between singlet and octet scalar resonances,
the asymptotic short-distance behaviour of the scalar form-factors and
some preliminary results on  $L_6(\mu)$.

\vspace*{-0.25cm}
\section{Octet $SS-PP$ correlator}

The two-point Green function $SS-PP$ is defined as
\begin{eqnarray}
&&
\hspace*{-1.4cm}
\Pi_{S-P}^{ab}(p) =
i \hspace*{-0.1cm}\int \hspace*{-0.15cm}d^4x\,  e^{ip\cdot x}
\langle 0| T[S^a(x)S^b(0)-P^a(x)P^b(0)]|0\rangle
\nn\\&&
= \delta^{ab}\Pi(p^2)\,,
\end{eqnarray}
with the SU(3) octet densities  
$S^a=\bar{q}\frac{\lambda_a}{\sqrt{2}} q$ and
$P^a=i\bar{q}\frac{\lambda_a}{\sqrt{2}}\gamma_5 q$,
being $\lambda_a$ the Gellmann matrices ($a=1,\dots 8$).

In the chiral limit, assumed all along the paper, the low-energy
expansion of the   correlator in $\chi$PT up to $\cO(p^6)$
has the form~\cite{SS-PP-Bijnens},
\begin{eqnarray}
&& \hspace*{-1.4cm}
\Frac{1}{B_0^2}\Pi(p^2)  = \frac{2F^2}{p^2}
+ \left[ 32L^r_8(\mu_\chi)
+\frac{\Gamma_8}{\pi^2}\left(1-\ln\frac{-p^2}{\mu_\chi^2}\right)
\right]
\nn
\\&&
\hspace*{-1cm}
+\frac{p^2}{F^2}\left[32C_{38}^r(\mu_\chi)
-\frac{\Gamma^{(L)}_{38}}{\pi^2}\left(1-\ln\frac{-p^2}{\mu_\chi^2}\right)
+ {\cal O}(N_C^0)\right] \, ,
\label{eq.SS-PP-chpt}
\end{eqnarray}
where   $\Gamma_8=5/48$ [$3/16$] and $\Gamma^{L}_{38} = -5L_5/6$ [$-3 L_5/2$]
in $SU(3)$--$\chi$PT  [$U(3)$--$\chi$PT]~\cite{chpt,Leutwyler-Kaiser}.
The coupling $L_8$ is quite relevant in
quantum chromodynamics (QCD)  phenomenology,
as it is one of the $\cO(p^4)$ LEC that rule the relation
between the quark and the pseudo-Goldstone masses~\cite{chpt}.

Within the large--$N_C$ limit, the amplitudes is given by
tree-level meson exchanges:
\begin{equation}
\hspace*{-0.7cm}
\Frac{1}{B_0^2} \Pi (p^2) =  \frac{2 F^2}{p^2}+ 16 \sum_{i}\left(
\frac{c_{m,i}^2}{M_{S,i}^2-p^2}- \frac{d_{m,i}^2}{ M_{P,i}^2-p^2}\right)
\, ,
\label{LO}
\end{equation}
where the sum goes   over the different  resonance multiplets.

In general, there is a reduced knowledge on the spectrum at high energies
and just the lightest resonances  are moderately well known.
Thus, one is forced in many cases to truncate the infinite tower
of large--$N_C$ states and to work with a finite number of them or even
within the single resonance approximation, as we did  in the
present work~\cite{L8-Trnka}.
Still, one may think about interpolating this rational approximation between
low-energies --ruled by $\chi$PT-- and the deep Euclidean domain
at $p^2\to -\infty$ --governed by the OPE--~\cite{Masjuan-Pade}.
After demanding
for $\Pi(s)$ the short-distance OPE
behaviour~\cite{RChTb,SVZ,Weinberg2,MHA},
this minimal hadronical  approximation  produces the
large--$N_C$ Weinberg sum-rules (WSR)~\cite{PI:08}
\begin{equation}
\hspace*{-0.6cm}
2F^2+ 16 d_m^2- 16 c_m^2    =0  , \,\,\,
 16 d_m^2 M_P^2 - 16 c_m^2 M_S^2   =\epsilon  ,
\label{eq.LO-WSR}
\end{equation}
where the tiny contribution $\epsilon$ from the
dimension four condensate~\cite{conde-O4}
is usually neglected, as we do in the present work~\cite{L8-Trnka}.

However, there are several caveats that may be raised to
this kind of procedures since
the truncation might, in principle,
induce important uncertainties in the LEC
determinations~\cite{Golterman-L8}.
Indeed, if one analyzes a growing number of observables
eventually one reaches inconsistences between high-energy constraints
from different amplitudes~\cite{Prades-GF,L10-Pich}.
Thus, the meson couplings of the truncated theory cannot be the same as in full QCD,
being closer for the lightest states but, possibly,
very different for the heaviest ones~\cite{Masjuan-Pade}.  
The conclusion  is that there is always a price to pay when applying
high-energy constraints to truncated large--$N_C$ 
amplitudes~\cite{Cillero-alphas}.

But the price can be paid in different ways and one can check
out the approximations  
by observing how much the outcomes differ
between one approach and another, e.g. between
the determination from Ref.~\cite{L8-Pich} and this work~\cite{L8-Trnka}.
Thus, the $SS-PP$ dispersive calculation
carried out in Ref.~\cite{L8-Pich} was repeated here
by explicitly computing the different diagrammatic contributions
up to one loop~\cite{L8-Trnka}.
The main difference was that now we demanded high-energy constraints
for the whole correlator $\Pi(s)$ and not for each separate
two-meson cut.  Likewise, We did not consider short-distance conditions
from other amplitudes, as it was done in Ref.~\cite{L8-Pich},
where constraints from the $VV-AA$ correlator and the vector and scalar
form-factors were taken into account.

The starting point of all the analysis is the LO Lagrangian, 
which in addition
to the leading tree-level amplitudes (LO in $1/N_C$),
also generates the leading one-loop diagrams  (next-to-leading order
in $1/N_C$).
The full large--$N_C$ generating functional is then approached
by the addition of more and more complicated terms to the
R$\chi$T action.  First we consider  the simplest operators
$\mL_G$ and $\mL_R$, with, respectively, only Goldstones and
one resonance field together with any number of Goldstones~\cite{RChTa}.  
This is the relevant Lagrangian
if one remains at tree-level. But at one-loop one needs to provide also a good
description of the cuts with one resonance and one chiral Goldstone.
One has then to consider  the next relevant operators,
\begin{eqnarray}
&& \hspace*{-1cm}
\mL_{RR'}= i \lambda_1^{PV} \bra [\nabla^\mu P,V_{\mu\nu}]  u^\nu\ket
+\lambda_1^{SA} \bra \{\nabla^\mu S,A_{\mu\nu}\}  u^\nu\ket
\nn\\
&&\qquad
+  \lambda_1^{SP} \bra \{ \nabla^\mu S,P\}   u_\mu\ket\, ,
\label{eq.LRR}
\end{eqnarray}
which contain  Goldstones and
two-resonance fields (as one of them can be in
the $s$--channel connected with the external scalar 
and pseudoscalar sources).
In principle, one could go further and include also operators
with three resonance fields to mend the  
two-resonance cuts. Nonetheless, we already found a good convergence
to our final result without them. The $RR'$ channels have
their thresholds at $(M_R+M_{R'})\sim 2$~GeV and are
suppressed enough at low energies 
for our current level of precision~\cite{L10-Pich}.

In general, at next-to-leading order in $1/N_C$ (NLO)
the perturbative calculation of the
$SS-PP$ correlator $\Pi(p^2)$  shows  the generic structure~\cite{L8-Trnka}
\begin{eqnarray}
&&\hspace*{-1.5cm}\Frac{1}{B_0^2}\, \Pi (p^2)=
\label{eq.NLO-correl2}
\\
&&\hspace*{-1.5cm}\quad \Frac{16 c_m^{ 2}}{M_S^{  2} -p^2}
- \Frac{16 d_m^{ 2}}{M_P^{ 2} -p^2}
+\, \Frac{2 F^2}{p^2}
+ 32 \widetilde{L}_8  +\Pi^{r}(p^2)^{1-\ell oop}\, ,
\nn
\end{eqnarray}
where $\Pi^{r}(p^2)^{1-\ell oop}$ contains the finite part of the
one-loop diagrams after renormalizing the couplings $c_m$, $d_m$, $M_S$, $M_P$
and $\widetilde{L}_8$. Notice that  other 
possible NLO operators of the R$\chi$T Lagrangian
that could contribute  to $\Pi(s)$
are proportional to the equations of motion and
have been already removed from the action
for the computation of Eq.~(\ref{eq.NLO-correl2}),
by means of  convenient meson field
redefinitions~\cite{L8-Trnka,Cillero-thesis,Rosell-thesis,L9-Pich}.
Their information is then encoded in the remaining  surviving
couplings $c_m$, $d_m$, $M_S$, $M_P$
and $\widetilde{L}_8$.

Next, we demand (\ref{eq.NLO-correl2}) to vanish like $1/p^6$
at high Euclidean momentum, as OPE dictates
(after neglecting the tiny dimension four
condensate~\cite{conde-O4,PI:08}).   At the one-loop level one has
\begin{eqnarray}
\Pi(p^2)=\sum_{k=0, 1, 2...} \Frac{1}{(p^2)^k}\left(
\alpha_{2k}^{(p)}+\alpha_{2k}^{(\ell)}\ln\Frac{-p^2}{\mu^2}\right)\, ,
\end{eqnarray}
where now  there are  $\frac{1}{(p^{2})^n}\ln(-p^2)$ 
terms with $n=0,1,2$ which provide
three large--$N_C$ conditions 
$\alpha_{0}^{(\ell)}=\alpha_{2}^{(\ell)}=\alpha_{4}^{(\ell)}=0$.  From
the non-log constraints
we find, on one hand, the value $\widetilde{L}_8(\mu)=0$~\cite{NLO-satura} 
and, on the other, the WSRs of~(\ref{eq.LO-WSR}) get now
NLO corrections $A(\mu)$ and $B(\mu)$,
\begin{eqnarray}
&& \hspace*{-1cm}
 2F^2+ 16 d_m(\mu)^2- 16 c_m(\mu)^2  + A(\mu)=0   , \,\,\,
\nn\\
&&  \hspace*{-1cm}
16 d_m(\mu)^2 M_P(\mu)^2 - 16 c_m(\mu)^2 M_S(\mu)^2 +B(\mu) =0  .
\label{eq.NLO-WSR}
\end{eqnarray}
These relations  allow us to fix  the renormalized couplings
$c_m(\mu)$ and $d_m(\mu)$ up to NLO in $1/N_C$~\cite{L8-Trnka,L8-Pich}.

Now, we are in conditions to  perform
the low-energy limit. The first thing to notice
is that one recovers exactly
the coefficients of the chiral logs from $\chi$PT,
independently of the value of the
R$\chi$T couplings. Thus, the matching with $\chi$PT is always possible
and the running of the LEC can be always recovered~\cite{NLO-satura},
producing here the LEC predictions~\cite{L8-Trnka}
\begin{eqnarray}
L_8(\mu ) &=&  \frac{ c_m(\mu)^2}{2\,M_S(\mu)^2}
- \frac{ d_m(\mu)^2}{2\, M_P^2}
+  \widetilde{L}_8(\mu) +  \xi_{L_8}(\mu) \, ,
\nn\\
C_{38}(\mu )&=& \frac{ F^2 c_m(\mu)^2}{2\, M_S(\mu)^4}
-\frac{ F^2 d_m(\mu)^2}{2\,M_P(\mu)^4}
+ \xi_{C_{38}}(\mu)
\, .
\end{eqnarray}
The leading terms, with $c_m^2$ and $d_m^2$, come from the
low-energy expansion of the tree-level resonance exchanges,
$\widetilde{L}_8$ from the local R$\chi$T contribution
and $\xi_{L_8, C_{38}}$ from
the low-energy expansion of renormalized
one-loop diagrams. The last simplification comes
after substituting  the   high-energy OPE
constraints~\cite{SVZ,Weinberg2,RChTb}, which set   
$\widetilde{L}_8(\mu)=0$~\cite{L8-Trnka,NLO-satura},
fix $c_m(\mu)$ and $d_m(\mu)$
through the NLO WSR~(\ref{eq.NLO-WSR}),
and imposes the three logarithmic
large--$N_C$ constraints 
$\alpha_0^{(\ell)}=\alpha_2^{(\ell)}=\alpha_4^{(\ell)}=0$
previously referred~\cite{L8-Trnka}.

If we now have a look at the phenomenology and make a first large--$N_C$
estimate, we get~\cite{L8-Trnka}
\begin{eqnarray}
L_8=(0.8\pm 0.3) \cdot 10^{-3}\, ,\,\,\,
C_{38}=(8\pm 5) \cdot  10^{-6}\, ,
\end{eqnarray}
with $M_S=980\pm 20$~MeV, $M_P=1300\pm 50$~MeV, $F=90\pm 2$~MeV and
the error given essentially by the naive error in the saturation scale
$\mu$, which was varied between $0.5$ and $1$~GeV.

At  NLO in $1/N_C$  with the simplest Lagrangian
with operators $\mL_G+\mL_R$, with at 
most one resonance field~\cite{RChTa},       
the predictions go completely off, yielding
$L_8(\mu_0)=(2.28\pm 0.19) 10^{-3}$ and
$C_{38}(\mu_0)=(26\pm 4) 10^{-6}$
for   the standard comparison scale
$\mu_0=770$~MeV,  really far away from the
usual  determinations in the
bibliography~\cite{chpt,L8-Pich,L8-others}.
The predictions immediately move towards  the right direction
after including any of the three two-resonance operators $\mL_{RR'}$
of~(\ref{eq.LRR}). These three new determinations converge
into the values
$L_8(\mu_0)=(1.1\pm 0.3) 10^{-3}$ and
$C_{38}(\mu_0)=(9\pm 4) 10^{-6}$
if one includes   the $\mL_{PV}$ and $\mL_{SP}$
operators from~(\ref{eq.LRR}), which are essential for the description
of the lightest thresholds ($V\pi$ and $S\pi$)  beyond the $\pi\pi$ one.
No relevant variation is found after adding to this the last
operator $\mL_{SA}$ in Eq.~(\ref{eq.LRR}), which is needed for the
$A\pi$ channel. Thus, we obtain our final outcome~\cite{L8-Trnka},
\begin{equation}
\hspace*{-0.7cm}
L_8(\mu_0)=(1.0\pm 0.4) \cdot  10^{-3}\, ,
\,\,\,  C_{38}(\mu_0)=(8\pm 5) \cdot   10^{-6}\, ,
\end{equation}
in relatively good agreement with former
determinations~\cite{chpt,L8-Pich,L8-others}.

\vspace*{-0.25cm}
\section{$S_1S_1-S_8S_8$ correlator}

We will have a look now at the connected correlator of two scalar densities
of different flavour~\cite{L6-Moussallam}. In the chiral limit,
this correlator can be written as the difference of the SU(3) singlet
and octet scalar correlators
$\Pi_{1-8}(p^2)  \equiv
\Pi_{S_1 S_1}(p^2)  -  \Pi_{S_8S_8}(p^2) $, being provided
at $\cO(p^4)$ in $\chi$PT at low energies by~\cite{L6-Moussallam}
\begin{eqnarray}
&&
\hspace*{-1.cm}
\Frac{1}{B_0^2}\Pi_{1-8}(p^2)= 96  L_6(\mu_\chi)
+  \Frac{3  \Gamma_6}{\pi^2}
\left(1 - \ln{\Frac{-p^2}{\mu_\chi^2}} \right)   \,  ,
\end{eqnarray}
with $\Gamma_6^{SU(3)}=\frac{11}{144}$ and  $\Gamma_6^{U(3)}=\frac{1}{16}$.

In general, for the R$\chi$T Lagrangians we considered 
(those from previous section),
we found in  all the allowed one loop diagrams
the singlet--octet relation,
\begin{equation}
\Pi_{S_1 S_1}(p^2)^{1-\ell oop}\,\,=\,\,
2\,\,\Pi_{S_8 S_8}(p^2)^{1-\ell oop}\, ,
\end{equation}
which means that the two--meson cut spectral function obeys
the positivity relation
\begin{equation}
\mbox{Im}\Pi_{1-8}(t)=\mbox{Im}\Pi_{S_8S_8}(t)
\geq 0\, .
\end{equation}
In addition, the OPE tells us that $\Pi_{1-8}(t)$ must vanish at short
distances like $1/t^2$~\cite{L6-Moussallam}.
Hence every separate two-meson contribution to Im$\Pi_{S_8S_8}(t)$
has to vanish like $1/t^2$ when $t\to\infty$.
The reason for this factor--2 relation is the symmetric structure
of the meson vertices. Thus, if  the flavour flow of
a general diagram in our R$\chi$T  calculation
is analyzed,  one explicitly finds this factor:
\begin{equation}
\hspace*{-0.5cm}
\bra T^a\{T^c,T^d\}\ket \, \bra \{T^c,T^d\}T^b\ket\,\,\,=\,\,\,
6\, (\delta^{ab}+\delta^{a0}\delta^{b0})\, .
\end{equation}
The octet generators would correspond to $T^a$ with $a=1, ... 8$ and
the SU(3) singlet one would be $T^0$.

This leads to the constraints obtained in previous
works for two-meson scalar 
form-factors~\cite{L8-Pich,L10-Pich,Rosell-thesis}:
$\Frac{4 c_d c_m}{F^2} =1$
($\pi\pi$ channel),  $\lambda_1^{SP} =-\Frac{d_m}{c_m}$
($P\pi$ channel),
$\lambda_1^{SA} =0$
($A\pi$ channel).

If we repeat the procedure applied to the $SS-PP$ correlator
$\Pi(s)$
and impose the OPE high-energy behaviour on $\Pi_{1-8}(s)$,
we obtain a pretty simple prediction for the
correlator at low energies and its corresponding LEC:
\begin{eqnarray}
\label{eq.L6}
&& \hspace*{-1.4cm}L_6(\mu )^{SU(3)} =
  \, \Frac{ c_m(\mu)^2}{6} \left(\Frac{1}{M_{S_1}^2(\mu) }
  -  \Frac{1}{M_{S_8}^2(\mu)}\right)
\\
&&\hspace*{-0.9cm}
\,\,  -\,\,\Frac{d_m^2}{32\pi^2 F^2}\left(1+\Frac{2 M_P^2}{M_S^2}\right)
\,\ln\Frac{M_P^2}{\mu^2}
\,\, + \,\, \Frac{1}{2304\pi^2 }\ln\Frac{m_{0}^2}{\mu^2}
\, ,
\nn
\end{eqnarray}
with the last term given by
$(\Gamma_6^{SU(3)}-\Gamma_6^{U(3)})/32\pi^2$
and the $\eta_1$ mass $m_0$,  and
providing the matching between U(3) and SU(3)
$\chi$PT~\cite{Leutwyler-Kaiser}.  The first bracket on the r.h.s. contains
the tree-level and the one-loop $\pi\pi$ contribution.
The $A\pi$ channel turns out to be  exactly zero after performing the
short-distance matching, which demands $\lambda_1^{SA}=0$.
The $P\pi$ loops contribute through the term proportional to $d_m^2$
in Eq.~(\ref{eq.L6}).

\vspace*{-0.35cm}
\section{Conclusions}

\vspace*{-0.20cm}
The different caveats about the high energy matching
and the truncation of the large--$N_C$
spectrum~\cite{Golterman-L8,Prades-GF,L10-Pich,Masjuan-Pade}
were confronted in Ref.~\cite{L8-Trnka}.
Although we worked with a resonance theory with a finite number of
mesons and terms in the Lagrangian, a nice convergence
to our final outcomes was found as more and more operators were 
added to the R$\chi$T action.
Our result for the octet $SS-PP$ correlator was  found
to be consistent with those from previous
analyses~\cite{chpt,L8-Pich,L8-others}.

Following the same procedure, we have started the analysis of
the difference $\Pi_{1-8}(s)$
between the SU(3) singlet and octet scalar correlators,
which also vanishes  at short distances~\cite{L6-Moussallam}  
and provides high-energy  constraints.  
The structure of the vertices has led to a positivity relation
for the two-meson spectral function. Hence,
each separate two--meson cut and the corresponding scalar form-factors
must vanish individually at
$p^2\to \infty$~\cite{L8-Pich,Rosell-thesis,L10-Pich}.
This allows one to fix most of the resonance parameters entering
in this amplitude
and yields a pretty compact expression for the corresponding
low-energy constant $L_6(\mu)$ in terms of R$\chi$T parameters.











\end{document}